\documentclass[a4paper,fleqn,10pt]{article}
 \usepackage[final]{graphicx}
\usepackage{graphics}
\usepackage[mathscr]{eucal}
\usepackage{euler}
\usepackage{amsmath}
\usepackage{amssymb}
\usepackage{subeqnarray}
\usepackage{graphicx} 
\usepackage{epsfig}

\begin{document}

\title{Limiting phase trajectories and the origin of energy localization in nonlinear oscillatory chains}

\author{L.I. Manevitch, V.V. Smirnov \footnote{vvs@polymer.chph.ras.ru}\\
\small{Institute of Chemical Physics, RAS}\\ \small{119991, 4 Kosygin str., Moscow, Russia}}

\date{June 7, 2010}

\maketitle

\begin{abstract}
We demonstrate that the modulation instability of the zone boundary mode in a finite (periodic) Fermi-Pasta-Ulam chain is the necessary but not sufficient condition for the efficient energy transfer by localized excitations. This transfer results from the exclusion of complete energy exchange between spatially different parts of the chain, and the excitation level corresponding to that turns out to be twice more than threshold of zone boundary mode's instability. To obtain this result one needs in far going extension of the beating concept to a wide class of finite oscillatory chains. In turn, such an extension leads to description of energy exchange and transition to energy localization and transfer in terms of \textit{'effective particles'} and \textit{Limiting Phase Trajectories}.  The 'effective particles' appear naturally when the frequency spectrum crowding ensures the resonance interaction between zone boundary and two nearby nonlinear normal modes, but there are no additional resonances. We show that the Limiting Phase Trajectories corresponding to the most intensive energy exchange between 'effective particles' can be considered as an alternative to Nonlinear Normal Modes, which describe the stationary process.
\end{abstract}

\section{Introduction}
It is well known that existence of spatially localized excitations is considerably important in different fields of nonlinear physics. Excitations of this type determine elementary mechanisms of many physical processes and make a noticeable contribution into thermal capacity \cite{r1}. Over the last decades spatially localized excitations were studied in the frameworks of two fields of nonlinear dynamics, namely, the theory of solitons \cite{r2}-\cite{r5} and the theory of nonlinear normal modes (NNMs) \cite{r6,r7,r8}. The theory of solitons was predominantly developed in application to the infinite continuous models described by nonlinear partial differential equations. As for NNMs, they can be considered as an extension of linear normal modes (LNMs) to discrete nonlinear systems. Development of both theories has revealed the soliton excitations in discrete systems (Toda \cite{r5} and Ablovitz-Ladik \cite{r9} lattices) as well as NNMs in continuum models \cite{r8}. Existence of the envelope solitons (breathers) allows us to find a common point for both theories \cite{r8}.

One of the widely used models of Nonlinear Dynamics is Fermi-Pasta-Ulam (FPU) \cite{r10} oscillatory chain study of which has become a starting point for discovery of solitons \cite{r11}. Long- and short-wavelength approximations of the FPU-chain are integrable Korteweg de Vries (KdV) and Nonlinear Schr\"odinger (NSE) equations, respectively \cite{r1, r12}. These equations in the limit of infinite chain have dynamical solitons (KdV) and envelope solitons (NSE) as their particular solutions. As for the finite FPU chains, their nonlinear dynamics has been intensively studied in terms of NNMs \cite{r13} - \cite{r20}. 

These studies have shown that for low excitation level a slow energy transfer along such chains is realized by spatially periodic waves due to nonlinear resonant interaction between the degenerate NNMs corresponding to the same eigenvalue \cite{r19, r191, r20}. The detailed studies of zone-boundary mode ($\pi$-mode)  in the large enough systems have shown its instability above certain level of excitation \cite{r18, r21, r22,r222}. Such instability leads to spatial concentration of energy revealed in the form of chaotic breathers, which have been studied numerically \cite{r18}. However, a simplest model of two weakly coupled nonlinear oscillators (2DoF) allows one to suppose that not one but two dynamical transitions can occur in the systems under consideration \cite{r23}. In the case of 2DoF system the first of the transitions is caused by instability of the out-of phase mode. It is important to note that a complete energy exchange between the oscillators can occur both above and below this threshold. But a second threshold exists for a higher excitation level and after its crossing the complete energy exchange is forbidden and the energy is confined in the initially excited oscillator \cite{r23}. The origin of second transition becomes clear if we consider the system's phase plane with using appropriate coordinates in which NNMs are depicted as stationary points. The loss of stability of out-of-phase NNM leads to creation of two new NNMs corresponding to weak energy localization in one of the oscillators. The domains of attraction of these two points are bounded by the separatrix passing through the unstable stationary point, and the trajectories providing the energy exchange are outside of separatrix. The trajectory corresponding to complete energy exchange is termed as Limiting Phase Trajectory (LPT) \cite{r23}. It has two branches which limit the domains of attraction of unstable (out-of-phase NNM) and stable (in-phase NNM) stationary points, respectively. While the excitation energy grows the domain of weak localization expands and the area between separatrix and the LPT diminishes up to complete degeneration of latter (when the separatrix coincides with the LPT). At this instant the complete energy exchange is forbidden and the energy becomes confined in the initially excited oscillator. Let us note that all these results could be naturally formulated in terms of the particles (oscillators) rather than of NNMs.

In the paper we would like to show that similar behavior is common for the nonlinear systems with the resonant interaction of NNMs and the analysis based on the LPT concept is a proper tool to clarify the fundamental problem of transition from energy exchange to energy localization in a finite (periodic) FPU chain. For this, one needs in far going extension of beating concept to the finite oscillatory chains. As to best our knowledge such an extension has not been elaborated till now. We have shown that introducing the LPT concept leads naturally to the notion of "effective particles" consisting of more than one real particles. Then the energy exchange in finite oscillatory chain can be described as beats in the system of two weakly coupled "effective particles" similarly to those for two real particles in 2DoF system.  

Our second goal is to fill a gap between the understanding of the origin of the vibration energy localization in two-degree-of-freedom and infinite systems. Namely, we derive the conditions of the transition to localization in a finite periodic FPU chain with more than two particles. It is worth mentioning that the FPU model can be mapped approximately to a DNLS (discrete Ablowitz-Ladik equation) with norm conservation  \cite{r24x, r24xx}. This result is significant for study of strongly discrete high-energy breathers because the Ablowitz-Ladik model is integrable one \cite{r9}. While dealing with such breathers there is no noticeable difference between infinite and finite chain \cite{r24, r24f, r24f2}. However, for considered low-energy excitations the difference is essential: only for finite chains both energy thresholds studied below exist. It is necessary to note also that the mechanism of energy localization considered in the paper differs from the scenario 'delocalization-localization' in the harmonic chains with embedded nonlinear oscillator \cite{r25}.

\section{The model}

We consider the finite $\beta$-FPU system with periodic boundary conditions defined by the Hamiltonian

\begin{equation} \label{GrindEQ__1_}
 H_{0} =\sum _{j}^{N}\frac{1}{2} p_{j} ^{2} +\frac{1}{2} (q_{j+1} -q_{j} )^{2} +\frac{\beta }{4} (q_{j+1} -q_{j} )^{4}   
\end{equation} 

where $q_j$ and $p_j$ are the coordinates and the conjugate momenta, respectively, and ($q_{N+1}=q_1$, $p_{N+1}=p_1$); \textit{N} is the quantity  of the particles. 

The transformation to the normal modes representation (Appendix A) leads to the equations of motion in the following form:

\begin{equation} \label{GrindEQ__6_}
\frac{d^{2} \xi _{k} }{dt^{2} } +\omega _{k} ^{2} \xi _{k} +\frac{\beta }{2N} \omega _{k} \sum _{l,m,n=1}^{N-1}\omega _{l} \omega _{m} \omega _{n} C_{k,l,m,n} \xi _{l} \xi _{m} \xi _{n}  =0.
\end{equation}

Here $\xi_k$ are the amplitudes of NNMs and the eigenvalues $\omega_k$ take the form

\begin{equation} \label{GrindEQ__4_}
\omega _{k} =\Omega\sin (\frac{\pi k}{N} ),\quad k=0,\ldots ,N-1.
\end{equation}

If considering an even number of particles $N$, the frequencies $\omega_k$ are bounded by the uppermost value $\omega_{N/2}=\Omega=2$. The lowermost eigenvalue $\omega_0=0$ corresponds to motion of the chain as a rigid body. All eigenvalues in the interval $0<\omega_k<2$ are doubly degenerate. If the number of the particles increases, the frequency gap between the highest-frequency mode and nearby modes quickly decreases with increase of $N$. Taking into account the dependence of the chain properties on the quantity of the particles, one can consider the value $\varepsilon=1/N$ as a small parameter of the system.

To analyze the dynamics of the chain in the context of Eqs \eqref{GrindEQ__6_}, we introduce the complex functions

\begin{equation} \label{_X_}
\Psi _{k} =\frac{1}{\sqrt{2} } (\frac{d\xi_k}{dt} +i\omega _{k} \xi _{k} ),\quad \Psi _{k}^{*} =\frac{1}{\sqrt{2} } (\frac{d\xi_k}{dt} -i\omega _{k} \xi _{k} ).
\end{equation} 

The complex functions $\Psi _{k}$ with indices ($k,~N-k$) correspond to the pairs of the degenerate modes and as it was mentioned above, their interaction provides the energy transfer by periodic traveling waves along the chain. A specific time scale for such energy transfer is proportional $\varepsilon^2$. Both the amount of the transferred energy  and the rate of the transfer depend on the value of the 'angular momentum' of the conjugate normal modes \cite{r19, r20}:

\[
G_{k} \sim i(\Psi _{k} \Psi _{N-k}^{*} -\Psi _{k}^{*} \Psi _{N-k} )\sim \omega _{k} (\xi _{N-k} \eta _{k} -\xi _{k} \eta _{N-k} ),
\]

The small value of $|G_k|$ ~corresponds to slow energy transfer, but the large amount of energy transferred, and vice versa. The limiting value of $G_k=0$ corresponds to zero rate of energy exchange when the energy transfer is absent. 

The situation changes drastically when the gap between the neighbor eigenvalues diminishes and there arise resonance relations between the normal modes corresponding to different eigenvalues. Actually, when the number of particles in the chain increases, the frequency gap between the $\pi$-mode and the nearest ones decreases as $\varepsilon^2$. Then a linear combination of the $\pi$-mode with the nearby conjugate ones may be considered as a long-wavelength modulation of $\pi$-mode with the ``wave number'' $1/N$.

In what follows, we consider the coupled dynamics of the $\pi$-mode ($k=N/2$) and the nearby conjugate modes with  indexes ($N/2+1$), ($N/2-1$) and linear frequency $\omega_{N/2-1}=\omega_{N/2+1}$. First of all it is necessary to note that the difference of frequencies of the $\pi$-mode and the nearby ones diminishes quickly while the number of particle grows:

\begin{equation} \label{M4}
\omega_{\frac{N}{2}\pm1}=\Omega\sin\frac{\pi (\frac{N}{2}-1)}{N}\approx\Omega(1-\frac{\pi^2}{2 N^2})=\omega_{N/2}(1-\frac{\pi^2}{2}\varepsilon^2)
\end{equation}

Therefore we choose $\omega_{N/2}$ as a basic frequency and take into account the resonant relationship between three highest-frequency modes. The multiple-scale analysis of equations of motion (see Appendix B for details) leads to  coupled equations \eqref{_B6_}, \eqref{_B7_}.

Keeping in mind the problem of energy exchange and transfer we may restrict ourselves by the following equations for three high-frequency modes:

\begin{equation} \label{GrindEQ__11_}
\begin{array}{l}
 {i\frac{d\chi _{N/2} }{d\tau _{2} } +\frac{3\beta }{4} [(|\chi _{N/2-1} |^{2} +|\chi _{N/2+1} |^{2} )\chi _{N/2} +(\chi _{N/2-1}^{2} +\chi _{N/2+1}^{2} )\chi _{N/2}^{*} ]=0} \\
  {i\frac{d\chi _{N/2-1} }{d\tau _{2} } -\frac{\pi ^{2} }{2} \chi _{N/2-1} +\frac{3\beta }{8} [(2|\chi _{N/2} |^{2} +|\chi _{N/2-1} |^{2} )\chi _{N/2-1} + (2\chi _{N/2}^{2} +\chi _{N/2+1}^{2} )\chi _{N/2-1}^{*} ]=0} \\
   {i\frac{d\chi _{N/2+1} }{d\tau _{2} } -\frac{\pi ^{2} }{2} \chi _{N/2+1} +\frac{3\beta }{8} [(2|\chi _{N/2} |^{2} +|\chi _{N/2+1} |^{2} )\chi _{N/2+1} + (2\chi _{N/2}^{2} +\chi _{N/2-1}^{2} )\chi _{N/2+1}^{*} ]=0} \end{array}
\end{equation}

There are three reasons for such restriction: (i) if the low-frequency modes are not exciting from very beginning they have no effect on the behavior of high-frequency modes; (ii) the high-frequency modes are dominant ones if the initial excitation is spatially localized; (iii) our computer simulation data show that the effect of low-frequency constituents is inessential even if their amplitudes are close to those of high-frequency modes.

The eqs \eqref{GrindEQ__11_} correspond to the Hamiltonian

\begin{equation} \label{GrindEQ__12_}
 \begin{array}{l} 
 {H=-\frac{\pi ^{2} }{2} (|\chi _{N/2+1} |^{2} +|\chi _{N/2-1} |^{2} )+\frac{3\beta }{8} [\frac{1}{2} (|\chi _{N/2+1} |^{2} +|\chi _{N/2-1} |^{2} )^{2} +\frac{1}{2} (\chi _{N/2+1} \chi _{N/2-1}^{*} -\chi _{N/2+1}^{*} \chi _{N/2-1} )^{2} } \\
  {\quad \quad \quad(\chi _{N/2+1} \chi _{N/2}^{*} +\chi _{N/2+1}^{*} \chi _{N/2} )^{2} + (\chi _{N/2-1} \chi _{N/2}^{*} +\chi _{N/2-1}^{*} \chi _{N/2} )^{2} ]} 
\end{array} 
\end{equation}

Eqs \eqref{GrindEQ__11_} possess an additional integral of motion, which is said to be the total 'occupation number':

\begin{equation} \label{GrindEQ__13_}
X=|\chi _{N/2} |^{2} +|\chi _{N/2-1} |^{2} +|\chi _{N/2+1} |^{2} =const
\end{equation}
(we omit the index of X because in following we will consider the highest-frequency modes only).
\section{The 'effective oscillators' description}
To analyze the system, we turn from the 'wave representation' to the 'particle representation'. Really, each of the modes considered represents a delocalized solution with the spatially uniform ($\pi$-mode) or periodic (conjugate modes) energy distribution. However, as it was mentioned above, a combination of the modes gives raise to a spatially non-uniform solution, corresponding to the energy localization in the one half of the chain. So, choosing a linear combination of the  considered modes we make obvious an analogy with the system of two nonlinear oscillators, where each half of the chain is considered as an 'effective oscillator'. This implies the transition from the 'wave' to 'particle' representation. In such a case the beating in the system of the 'effective oscillators' manifests as the process of periodic energy transfer from one half of the chain to another one and inversely with time scale $\varepsilon^2$ . To this end, we introduce the 'particle' representation through the variables $\psi_1$, $\psi_2$, $\varphi$, defined by the following formulas

\begin{equation} \label{GrindEQ__14_}
\begin{array}{l} {\psi _{1} =\frac{1}{\sqrt{2} } (\chi _{N/2} - (\sqrt{1-c^2} \chi _{N/2-1} + c \chi _{N/2+1} ))} \\ {\psi _{2} =\frac{1}{\sqrt{2} } (\chi _{N/2} + (\sqrt{1-c^2} \chi _{N/2-1} + c \chi _{N/2+1} ))} \\ {\varphi = (c \chi _{N/2-1} -\sqrt{1-c^2} \chi _{N/2+1} )} \end{array},
\end{equation}
where $c$ is a constant defined by initial conditions ($0\leq c \leq 1$). Such transformation preserves the occupation number in the form $X=|\psi_1|^2+|\psi_2|^2+|\varphi|^2$.

We consider first a particular solution corresponding to $\varphi=0$. Since the occupation number $X$ is the integral of motion, the variables $\psi_1$ and $\psi_2$ can be expressed through the 'angular variables' \cite{r26}:

\begin{equation} \label{GrindEQ__15_}
\psi _{1} =\sqrt{X} \cos \theta e^{i\delta _{1} } ;\quad \psi _{2} =\sqrt{X} \sin \theta e^{i\delta _{2} } . 
\end{equation}

Then the Hamilton function \eqref{GrindEQ__12_} takes the following form in the terms of 'angular variables':

\begin{equation} \label{GrindEQ__16_}
\begin{array}{l}
{H(\theta ,\Delta )=\frac{X}{64} [27\beta X-16\pi ^{2} +2(8\pi ^{2} -3\beta X)} \\ {\quad \times \cos \Delta \sin 2\theta -3\beta X(8-\cos ^{2} \Delta )\sin ^{2} 2\theta ]},
\end{array}
\end{equation}

where $\Delta=\delta_2-\delta_1$.

The analysis of phase-plane portrait of the Hamiltonian \eqref{GrindEQ__16_} shows that there exist two stationary points corresponding to the pure $\pi$-mode ($\theta=\pi/4$, $\Delta=0$) and  the``pure'' mixture of conjugate modes ($\theta=\pi/4$, $\Delta=\pi$), respectively, at a small value of the occupation number $X$ (Fig. 1a). If we consider the phase trajectory associated with the transition from the state $\psi_1$ ($\theta=0$) to the state $\psi_2$ ($\theta=\pi/2$), then we find that it corresponds to beating in the system of the ``effective oscillators''. Since this trajectory represents an outer boundary for a set of trajectories encircling the basic stationary points (Fig. 1a), we refer to it as the Limiting Phase Trajectory (LPT) \cite{r23}. The motion along this trajectory leads to the complete reversible energy transfer between two ``effective oscillators''.

\begin{figure}
	\noindent\centering{
	\includegraphics{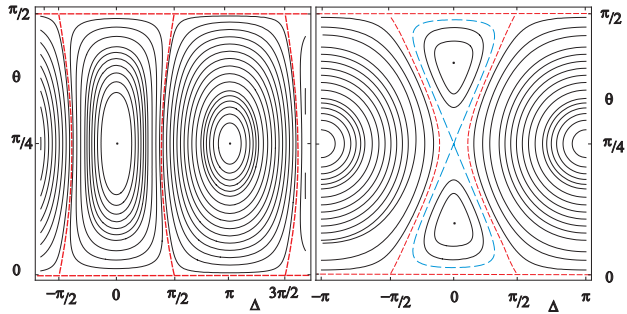}
	}
\caption{ (color online)\textbf{ }The truncated phase-plane portrait before (a) and after (b) threshold of instability of the p-mode; the (red) dashed and (blue) long-dashed curves correspond to the LPT encircling $\pi$-mode and the separatrix, respectively.}
\end{figure}

If the excitation level grows and attains a critical level the $\pi$-mode becomes unstable that leads to the appearance of two new stationary points (Fig. 1.b) with the folliwing coordinates :

\begin{equation}\label{GrindEQ__17_}
\begin{array}{l} {(a)\quad \Delta =0,\quad \theta =\frac{1}{2} \arcsin (\frac{8\pi ^{2} -3\beta X}{21\beta X} )} \\ {(b)\quad \Delta =0,\quad \theta =\frac{\pi }{2} -\frac{1}{2} \arcsin (\frac{8\pi ^{2} -3\beta X}{21\beta X} )} \end{array}.
\end{equation}

The threshold of instability of the $\pi$-mode is equal to $X_c=\pi^2/3\beta$, which is in a good accordance with the estimation obtained in the framework of the ``narrow packet'' approximation \cite{r22, r18}.

Any trajectories lying in the neighborhood of stationary points \eqref{GrindEQ__17_} correspond to weakly localized solutions, for which the energy of one half of the chain (``effective oscillator'')only slightly exceeds the energy of the second one. At the same time, the trajectories starting beyond this vicinity and passing through the points $\psi_1~(\theta=0)$ or $\psi_1~(\theta=\pi/2)$ correspond to the combinations of modes with approximately equal energies that means that an initial excitation of one ``effective oscillator'' entails the complete energy exchange between both ones, i.e. the transition from the state $\psi_1$ to the state $\psi_2$ and inversely. This implies that a possibility of complete energy exchange between different parts of the chain exists for the excitation level exceeding the instability threshold $X_c$.

A growth of the amplitude $X$ entails an enlargement of the domain encircled by the separatrix passing through an unstable stationary point (out-of-phase NNM); at last the  separatrix coincides  with the LPT.  At this instant topology of the phase plane changes drastically (Fig.2). Namely, any energy exchange between the ``mixed'' states $\psi_1$ and $\psi_2$ disappears; this implies that a trajectory starting at a point corresponding to $\theta<\pi/4$ (or $\theta>\pi/4$) and for any $\Delta$ cannot reach a point corresponding to $\theta>\pi/4$ ($\theta<\pi/4$) (excepting all the trajectories surrounding the in-phase stationary points $\Delta=\pm\pi$ and bounded by the separatrix crossing the unstable point ($\theta<\pi/4$, $\Delta=0$)). Therefore, the energy initially concentrated near the states $\psi_1$ or $\psi_2$ remains confined in the excited ``effective oscillator'' (Fig. 3).

\begin{figure}
\noindent
\centering{
\includegraphics{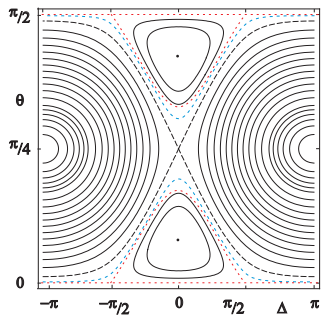}
}
\caption{(color online)\textbf{ }Phase-plane portrait after coincidence of the LPT and the separatrix; (black) long-dashed lines, and (blue) short-dashed lines correspond to the separatrix, passing through the unstable stationary point, and the transient trajectories, respectively; the (red) dotted line depicts the boundaries between the families of transient trajectories and closed ones encircling the stationary points.}
\end{figure}

An energy threshold associated with the above mentioned localization can be found from the condition of equality of the energy corresponding to the LPT and the energy at the unstable $\pi$-mode. It is seen in Fig. 1 that the LPT goes through the points $\theta=\pi/2$ and $\theta=0$. This means that

\[ H(\theta ,\Delta )_{|LPT} =\frac{X}{64} (27\beta X-16\pi ^{2} )\] 

Since the energy of the $\pi$-mode is equal to zero, it is easy to calculate the respective occupation number $X_{loc}=16\pi2/27\beta$ and the energy of the chain $X_{loc}=16\pi2/27\beta N$. It now follows that above the excitation level $E_{loc}$ we can observe the localized vibration excitation (a breather).

\begin{figure}
\noindent
\centering{
\includegraphics[width=70mm]{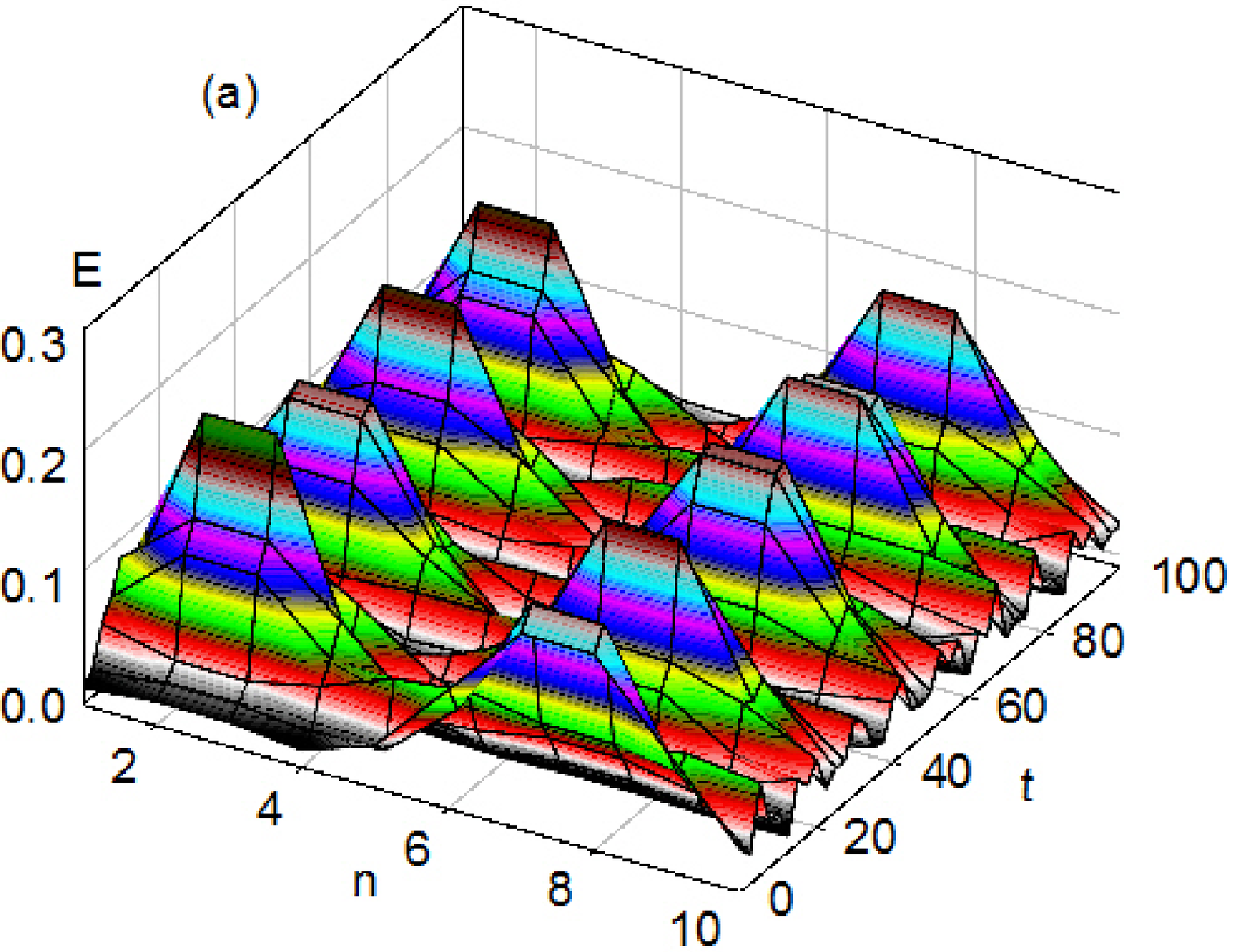}

\includegraphics[width=70mm]{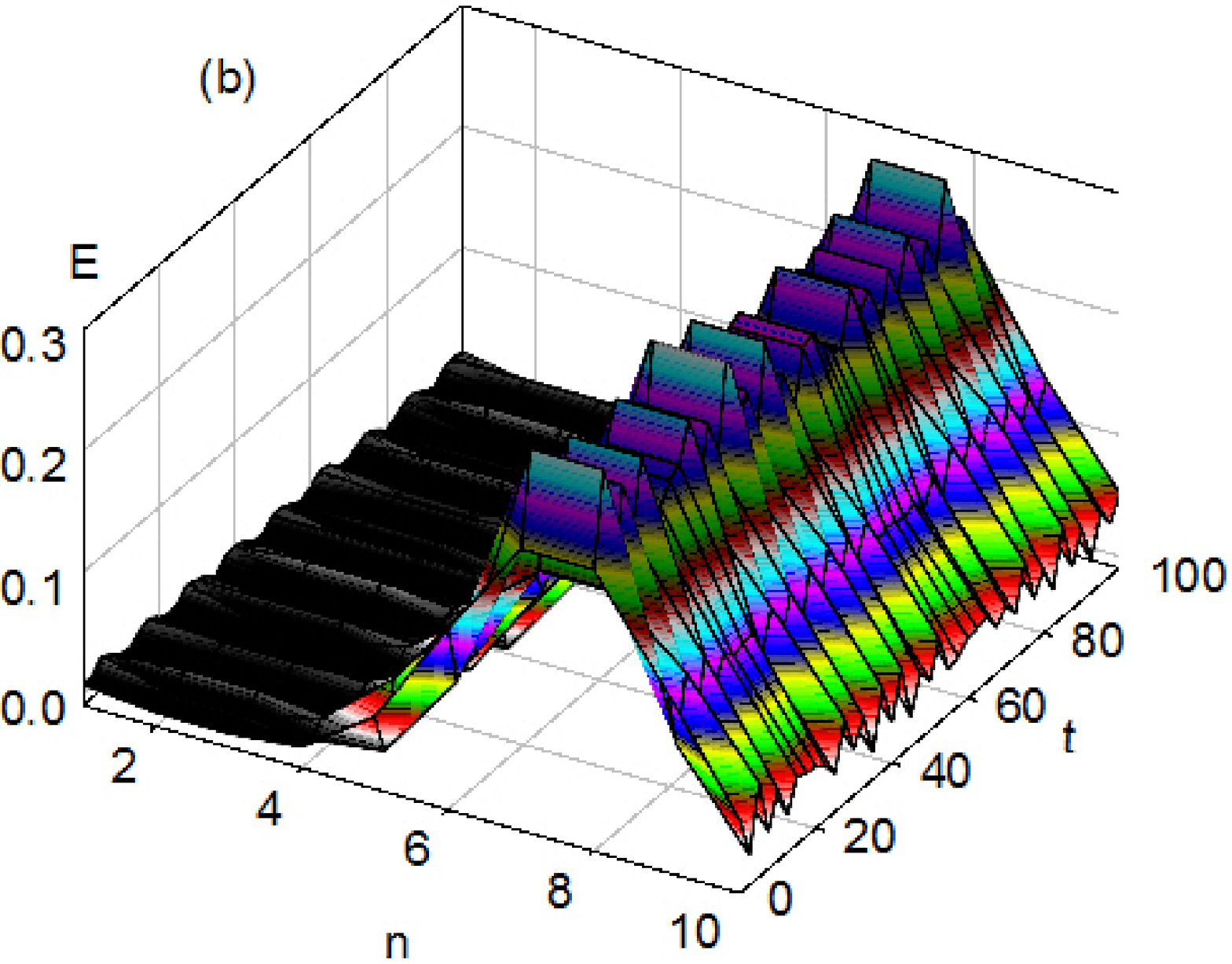}
}
\caption{(color online) Transition ``beating -- localization'' in the FPU-chain with 10 particles. The initial conditions correspond to the state $\psi_1$, and the levels of excitation are $X<X_{loc}$ (a) and $X>X_{loc}$ (b). $E$ and $n$ -- the energy (in the relative units) and the number of particles, respectively, $t$ -- the time (measured in the periods of $\pi$-mode - $T_0=2\pi/\omega_{N/2}$).}
\end{figure}

If the initial energy is concentrated at the state $\psi_1$ or $\psi_2$, the representing point in the phase plane moves along a trajectory encircling the respective stationary point. Then the temporal evolution of the breather corresponds to the regular variation of its profile (the ``breathing'' mode of the localized excitation).

The period of breathing can be calculated as

\[ T=\oint dt=\frac{1}{\varepsilon ^{2} } \oint \frac{d\Delta }{d\Delta /d\tau _{2} }   ,\] 

where the last integral is taken along the LPT.

Contrary to the ``breathing'' breathers corresponding to the motion along the LPT, the stationary points \eqref{GrindEQ__17_} determine new normal modes with an invariable non-homogeneous energy distribution, or the breather-like excitations. Although these modes exist at any $X>X_c$ the true threshold of localization is equal to $X_{loc}$ because of a possibility of the complete energy exchange is preserved for motion along the LPT till $X=X_{loc}$.

Thus we get the solution \eqref{GrindEQ__17_} corresponding to immobile breather. Finally we want to clarify the nature of traveling breathers. We recall that the above results have been obtained under assumption $\varphi =0$ that corresponds to equal amplitudes of the conjugate modes. Now we assume that the amplitudes of the conjugate modes are slightly different, that is we consider small enough $\varphi\neq 0$. One can show that the equations for functions $\psi_1$ and $\psi_2$ include only quadratic terms depending on $\varphi$. Therefore, in the framework of the linear approximation, if the value $\varphi\neq 0$ is small enough, there is no qualitatively change in the phase-plane portrait in figures (1, 2). The behavior of $\varphi$ in the vicinity of any of stationary points \eqref{GrindEQ__17_} is described by following equation:

\[
 i\frac{d\varphi }{d\tau _{2} } -\frac{\pi ^{2} }{2} \varphi +\frac{3}{16} \beta X[2\sin 2\theta \varphi +(3+\sin 2\theta e^{2i\delta _{1} } )\varphi ^{*} ]=0.
\] 

The respective eigenvalue

\[
\lambda =\pm \frac{\sqrt{891\beta ^{2} X^{2} +120\pi ^{2} \beta X-384\pi ^{4} } }{56} 
\] 

is imaginary if $X<X_{loc}$, i.e. if the amplitude of excitation is less than the threshold corresponding to the coincidence of LPT and the separatrix.

The distribution of the vibration amplitude along the chain can be characterized by the mean squared particle displacements $<q_j^2>$:

\[
<q_{j}^{2} >=\frac{\omega _{N/2} }{2\pi } \int _{0}^{2\pi /\omega _{N/2} }q_{j}^{2} (t,\tau _{2} )dt .
\] 

For $\varphi=0$, then there exists a single maximum of the vibration amplitude at $j=N/2$. It is easy to demonstrate that, if the eigenvalue $\lambda$ is imaginary, then a small nonzero $\varphi$  in the equations of motion leads to small oscillations of the breather center. At the same time, appearance of a real part in the eigenvalue $\lambda$ for $X>X_{loc}$ leads to the directional motion of the breather (Fig. 4).

\begin{figure}
	\noindent
	\centering{
	\includegraphics[width=70 mm] {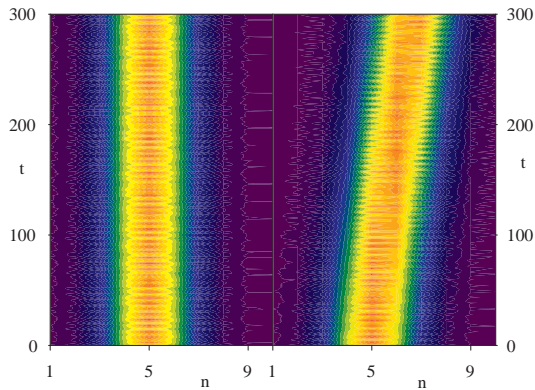}
	}
	\caption{(color online) Energy map for static ($X<X_{loc}$) (a) and traveling ($X>X_{loc}$) (b) breathers.}
\end{figure}

\section{Discussion}
Concluding, one can state that the energy localization in the finite nonlinear system like the FPU lattice is caused by the resonance interaction of nonlinear normal modes which are close to the spectrum boundary. At that, the adequate description of the process has achieved in the terms of ``effective oscillators'' rather than in terms of NNMs. There are three levels of the system excitation, at which the behavior of the system is obviously different. At a low excitation energy the resonance interaction of  modes leads to a phenomenon that is similar to beat oscillations in a system of two weakly coupled nonlinear oscillators, i.e. to periodic energy exchange between two ``effective oscillators''. Both the complete and partial energy exchange can be realized. The loss of stability of the zone-boundary mode leads to weak energy localization, when only a small excess of the energy is confined in the one ``effective oscillator''. Such energy excess increases with the growth of the excitation level. As this take place, the possibility of complete energy exchange is preserved in the attraction domain of ``out-of-phase'' oscillations, while both the complete and partial energy exchange can occur in the attraction domain of the ``in-phase'' oscillations. The center of the energy localization can oscillate under the influence of small perturbations. When the excitation level reaches the second threshold correlated with the degeneration of the area bounded by the LPT the beating converts into strong localization. This leads to the energy confinement in only one ``effective oscillators'' (i.e. in one part of the chain). It is important that this transition involves both ``out-of-phase'' and ``in-phase'' trajectories, i.e. the localization does not depend on the initial ``phase shift'' $\Delta$. A weak localization and non-complete energy exchange are still possible but their realization requires some special initial conditions. One of the key changes after the second transition is the transformation of small oscillations of the center of the localization area into a directed motion. Also, since the domain of strong localization does not correspond to any stationary point in the phase space of the system, the profile of the energy distribution along the chain demonstrates the periodic variations. A further growth of the excitation does not result in any qualitatively changes in the systems behavior, which might increase or decrease the areas corresponding to both partial energy exchange and weak localization. It should be note that increase of the particles number in the chain involves the new modes into resonance interaction with zone boundary mode because of spectrum crowding in the high-frequency range. It does not lead to any noticeable  changes except some "narrowing" of "effective oscillator". In the limiting case of infinite chain the "effective oscillator" transforms into well-known breather. At that, both energy thresholds discussed above tend to zero. It means that the resonant nature of energy localization can be better clarified while dealing with the finite chain.

\section{Conclusion}
Finally, we have revealed the origin of localization occurring due to resonance interaction in the range of the spectrum crowding in relatively small FPU chains. Using the LPT concept and the notion of "effective particles" has allowed us to clarify the origin of transition from energy exchange to energy localization and transfer in finite periodic oscillatory chains. We highlight the role of the size of systems under considerations. Since the gap between high-frequency modes decreases as $1/N^2$ the resonant relations are already realized even if the chain contains a small enough number of particles. Actually, the phenomenon of strong energy localization is seen clearly in the system with 8-10 particles, and the computer simulation data are in the excellent agreement with the analytic results. We would like to underline that the number of resonance relations increases and both discussed thresholds tend to zero when size of the chain grows. Therefore the origin of energy localization can be effectively treated if considering a finite chain with small enough number of resonances. Two non-zero thresholds exist for any finite chain and the only quantitative change in the case of large number of particles is decrease of the ``size'' of the ``effective particle'' because of the influence of new resonances. Small perturbations of the ``effective particles'' behavior owing to the modes, which can be ignored in the resonance approximation, become visible on very large time intervals (several thousands of the oscillation periods). We have observed that a long-run temporal evolution leads to the formation of localized excitations with the behavior similar to that of chaotic breathers.
\\

\textbf{Acknowledgments}

The work was supported by Program of Department of Chemistry and Material Science (Program \#1), Russia Academy of Sciences, and Russia Basic Research Foundation (grant 08-03-00420a).
\\
\\

\appendix{\bf\Large{Appendix}}

\section{The normal modes}

The finite $\beta$-FPU system with periodic boundary conditions described by the Hamiltonian \eqref{GrindEQ__1_} can be written in the terms of normal modes which are defined by the following linear canonical transformation

\begin{equation} \label{GrindEQ__a2_} q_{j} =\sum _{k=0}^{N-1}\sigma _{j,k} \xi _{k}^{}   \end{equation} 

where

\[\sigma _{j,k} =\frac{1}{\sqrt{N} } [\sin (\frac{2\pi kj}{N} )+\cos (\frac{2\pi kj}{N} )],\quad k=0,1,...,N-1. \] 

 Transformation \eqref{GrindEQ__a2_} allows us to present the quadratic part of the Hamiltonian as the total energy of the independent oscillators:

\begin{equation} \label{GrindEQ__a3_} 
H_{2} =\sum _{k=1}^{N-1}\frac{1}{2} (\eta _{k} ^{2} +\omega _{k}^{2} \xi _{k}^{2} ).
\end{equation}

Here $\xi_k$ and $\eta_k$ are the amplitudes and the momenta of NNMs; the coordinate $\xi_0$, associated with motion of the center of mass, is removed from \eqref{GrindEQ__a3_}. The eigenvalues $\omega_k$ are defined by eq. \eqref{GrindEQ__4_}.

If the number of particles $N$ is  even one, the frequencies $\omega_k$ are bounded by the highest-frequency $\omega_{N/2}=\Omega=2$. The zero eigenvalue $\omega_0=0$ corresponds to motion of the chain as a rigid body. All eigenvalues in the interval $0<\omega_k<2$ are doubly degenerate.  The density of eigenvalues $\omega_k$ is the essentially nonuniform function of $k$: the lowest values (small $k$) are separated by large gaps while the distances between highest frequencies ($k$ near the $N/2$) are small enough and they diminish quickly with the growth of number of particles.

The 4th-order part of the Hamiltonian \eqref{GrindEQ__1_} for the periodic $\beta$-FPU chain has the following form \cite{r13}:

\begin{equation} \label{GrindEQ__a5_}
H_{4} =\frac{\beta }{8N} \sum _{k,l,m,n=1}^{N-1}\omega _{k} \omega _{l} \omega _{m} \omega _{n} C_{k,l,m,n} \xi _{k} \xi _{l} \xi _{m} \xi _{n},
\end{equation}

where 

\begin{equation} \label{_A1_}
\begin{array}{l} {C_{k,l,m,n} =-\Delta _{k+l+m+n} +\Delta _{k+l-m-n} } \\ {\quad \quad +\Delta _{k-l+m-n} +\Delta _{k-l-m+n} } \\ {\Delta _{r} =\left\{\begin{array}{l} {(-1)^{r} ,\; if\quad r=mN,\; m\in {\rm Z} } \\ {0\quad \quad otherwise.} \end{array}\right. } \end{array}.
\end{equation} 

It now follows from \eqref{GrindEQ__a3_}, \eqref{GrindEQ__a5_} that the equations of motion can be written in form \eqref{GrindEQ__6_}.

To analyze the dynamics of the chain we introduce the complex functions 

\[\Psi _{k} =\frac{1}{\sqrt{2} } (\eta _{k} +i\omega _{k} \xi _{k} ),\quad \Psi _{k}^{*} =\frac{1}{\sqrt{2} } (\eta _{k} -i\omega _{k} \xi _{k} ). \] 

In terms of these variables Eqs \eqref{GrindEQ__6_} can be rewritten as 

\begin{equation} \label{GrindEQ__a7_}
\begin{array} {l} 
{i\frac{d\Psi _k }{dt} +\omega _k \Psi _k -\frac{\beta }{8N} \omega _k \sum _{l,m,n=1}^{N-1}C_{k,l,m,n}}\\
{\quad\quad\quad \times (\Psi _{l} -\Psi _l^* )(\Psi _m -\Psi _m^* )(\Psi _n -\Psi _n^* )=0}.
\end{array}
\end{equation}

\section{The multiple scale expansion}
To consider the processes on the time intervals greatly exceeding $1/\omega_k$ we employ the multiple scale procedure. We introduce the time scales:

\begin{equation} \label{_B1_}
\tau _{0} =t,\quad \tau _{1} =\varepsilon \tau _{0} ,\quad \tau _{2} =\varepsilon ^{2} \tau _{0} ,\dots ,
\end{equation}

where the ``fast'' time $\tau_0$ corresponds to the original time scale of the system, while the slow times $\tau_1$, $\tau_2$, etc. correspond to the slowly varying envelopes.

We construct an asymptotic representation of function $\Psi_k$ in the form 

\begin{equation} \label{_B2_}
\Psi _{k} =\sqrt{\varepsilon } (\varphi _{k,1} +\varepsilon \varphi _{k,2} +\varepsilon ^{2} \varphi _{k,3} +...).
\end{equation}

Taking into account the relations \eqref{_B1_} and \eqref{_B2_} we get from Eqs \eqref{GrindEQ__a7_} in the lowest-order by small parameter $\varepsilon$:

\begin{equation} \label{_B3_}
\varepsilon ^{1/2} :\quad \left\{\begin{array}{l} {i\frac{\partial \varphi _{\frac{N}{2},1} }{\partial \tau _{0} } +\omega _{\frac{N}{2}} \varphi _{\frac{N}{2},1} =0} \\ {i\frac{\partial \varphi _{\frac{N}{2}\pm 1,1} }{\partial \tau _{0} } +\omega _{\frac{N}{2}-1} \varphi _{\frac{N}{2}\pm 1,1} =0} \end{array}\right.. 
\end{equation}

Taking the solutions of Eqs \eqref{_B3_} in the form

\[
\left\{\begin{array}{l} {\varphi _{N/2,1} =\chi _{\frac{N}{2},1} e^{i\Omega \tau_0} } \\ {\varphi _{\frac{N}{2}\pm 1,1} =\chi _{\frac{N}{2}\pm 1,1} e^{i\Omega \tau_0} } \end{array}\right. 
\] 

we get the discrepancy in the second of eq. \eqref{_B3_} that is the order of $\varepsilon^2$ and it has to be taken into account in the approximation of next orders. The amplitudes $\chi_{\frac{N}{2},1}$, $\chi_{\frac{N}{2}\pm1,1}$ do not depend on the 'fast' time $\tau_0$.

\[
\varepsilon ^{3/2} :\quad \left\{\begin{array}{l} {i\frac{\partial \varphi _{\frac{N}{2},2} }{\partial \tau _{0} } +\omega _{\frac{N}{2}} \varphi _{\frac{N}{2},2} +i\frac{\partial \varphi _{\frac{N}{2},1} }{\partial \tau _{1} } =0} \\ {i\frac{\partial \varphi _{\frac{N}{2}\pm 1,2} }{\partial \tau _{0} } +\omega _{\frac{N}{2}-1} \varphi _{\frac{N}{2}\pm 1,2} +i\frac{\partial \varphi _{\frac{N}{2}\pm 1,1} }{\partial \tau _{1} } =0} \end{array}\right. .
\] 

Taking

\[
\left\{\begin{array}{l} {\varphi _{\frac{N}{2},2} =\chi _{\frac{N}{2},2} e^{i\Omega \tau_0} } \\ {\varphi _{\frac{N}{2}\pm 1,2} =\chi _{\frac{N}{2}\pm 1,2} e^{i\Omega \tau_0} } \end{array}\right. ,
\] 

we get:

\[
\left\{\begin{array}{l} {i\frac{\partial \chi _{\frac{N}{2},1} }{\partial \tau _{1} } =0} \\ {i\frac{\partial \chi _{\frac{N}{2}\pm 1,1} }{\partial \tau _{1} } =0} \end{array} \right.
 \] 
The difference between $\omega_{\frac{N}{2}}$ and $\omega_{\frac{N}{2}-1}$ is negligibly small in this order by the small parameter $\varepsilon$.

\[ 
\varepsilon ^{5/2} :\quad \left\{
\begin{array}{l}
 {i\frac{\partial \varphi _{\frac{N}{2},1} }{\partial \tau _{2} } +i\frac{\partial \varphi _{\frac{N}{2},2} }{\partial \tau _{1} } +i\frac{\partial \varphi _{\frac{N}{2},3} }{\partial \tau _{0} } +\omega _{\frac{N}{2}} \varphi _{\frac{N}{2},3} = } 
\\ {\frac{\beta \Omega}{8} \sum _{l,m,n=1}^{N-1}C_{\frac{N}{2},l,m,n} [\varphi _{l,1} \varphi _{m,1} \varphi _{n,1} -3\varphi _{l,1} \varphi _{m,1} \varphi _{n,1}^{*} +  3\varphi _{l,1} \varphi _{m,1}^{*} \varphi _{n,1}^{*} -\varphi _{l,1}^{*} \varphi _{m,1}^{*} \varphi _{n,1}^{*} ]} 
\\ {i\frac{\partial \varphi _{N/2\pm 1,1} }{\partial \tau _{2} } +i\frac{\partial \varphi _{\frac{N}{2}\pm 1,2} }{\partial \tau _{1} } +i\frac{\partial \varphi _{\frac{N}{2}\pm 1,3} }{\partial \tau _{0} } +\omega _{\frac{N}{2}-1} \varphi _{\frac{N}{2}\pm 1,3} -\frac{\pi ^{2} }{2} \Omega  \varphi _{\frac{N}{2}\pm 1,1} =}
 \\ {\frac{\beta \Omega}{8}  \sum _{l,m,n=1}^{N-1}C_{\frac{N}{2}\pm1,l,m,n} [\varphi _{l,1} \varphi _{m,1} \varphi _{n,1} -   3\varphi _{l,1} \varphi _{m,1} \varphi _{n,1}^{*} +3\varphi _{l,1} \varphi _{m,1}^{*} \varphi _{n,1}^{*} -\varphi _{l,1}^{*} \varphi _{m,1}^{*} \varphi _{n,1}^{*} ]}
\end{array} \right.
\] 

Taking

\[
\left\{\begin{array}{l} {\varphi _{\frac{N}{2},3} =\chi _{\frac{N}{2},3} e^{i\Omega \tau_0} } 
\\ {\varphi _{\frac{N}{2}\pm 1,3} =\chi _{\frac{N}{2}\pm 1,3} e^{i\Omega \tau_0} } \end{array}\right. ,
\] 

we get:

\begin{equation} \label{_B4_}
\begin{array}{l} {i\frac{\partial \chi _{\frac{N}{2},1} }{\partial \tau _{2} } +i\frac{\partial \chi _{\frac{N}{2},2} }{\partial \tau _{1} } =} 
\\ {\frac{\beta \Omega}{8}  \sum _{l,m,n=1}^{N-1}C_{\frac{N}{2},l,m,n} [\varphi _{l,1} \varphi _{m,1} \varphi _{n,1} -3\varphi _{l,1} \varphi _{m,1} \varphi _{n,1}^{*} +   3\varphi _{l,1} \varphi _{m,1}^{*} \varphi _{n,1}^{*} -\varphi _{l,1}^{*} \varphi _{m,1}^{*} \varphi _{n,1}^{*} ]e^{-i\Omega \tau _{0} } } 
\\ {i\frac{\partial \chi _{\frac{N}{2}\pm 1,1} }{\partial \tau _{2} } +i\frac{\partial \chi _{\frac{N}{2}\pm 1,2} }{\partial \tau _{1} } -\frac{\pi ^{2} }{2} \Omega \chi _{\frac{N}{2}\pm 1,1} = } 
 \\ {\frac{\beta \Omega}{8} \sum _{l,m,n=1}^{N-1}C_{\frac{N}{2}\pm 1,l,m,n} [\varphi _{l,1} \varphi _{m,1} \varphi _{n,1} -  3\varphi _{l,1} \varphi _{m,1} \varphi _{n,1}^{*} +3\varphi _{l,1} \varphi _{m,1}^{*} \varphi _{n,1}^{*} -\varphi _{l,1}^{*} \varphi _{m,1}^{*} \varphi _{n,1}^{*} ]e^{-i\Omega \tau _{0} } } \end{array}
\end{equation}

To remove the second-order constituents from the last equations one should integrate them with respect to the times $\tau_0$ and $\tau_1$. To do it we have to know the manner in which the variables $\varphi_k$ depend on the 'fast' time. We can perform the similar procedure for all variables $\varphi_k$. 

\[ 
\varepsilon ^{1/2} :\quad \left\{\begin{array}{l} {i\frac{\partial \varphi _{k,1} }{\partial \tau _{0} } +\omega _{k} \varphi _{k,1} =0} \\ {i\frac{\partial \varphi _{N-k,1} }{\partial \tau _{0} } +\omega _{k} \varphi _{N-k,1} =0} \end{array}\right. 
\] 

The obvious solutions of the last equations are:

\[
\left\{\begin{array}{l} {\varphi _{k,1} =\chi _{k,1} e^{i\omega_k \tau_0} } \\ {\varphi _{N-k,1} =\chi _{N-k,1} e^{i\omega_k \tau_0} } \end{array}\right. ,
\]

where the amplitudes $\chi_{k,1}$ and $\chi_{N-k,1}$ do not depend on the time $\tau_0$. In the next order of $\varepsilon$ we get the following relationships:

\[
\varepsilon ^{3/2} :\quad \left\{\begin{array}{l} {i\frac{\partial \varphi _{k,2} }{\partial \tau _{0} } +\omega _{k} \varphi _{k,2} +i\frac{\partial \varphi _{k,1} }{\partial \tau _{1} } =0} 
\\ {i\frac{\partial \varphi _{N-k,2} }{\partial \tau _{0} } +\omega _{k} \varphi _{N-k,2} +i\frac{\partial \varphi _{N-k,1} }{\partial \tau _{1} } =0} \end{array}\right. .
\] 

\[
\left\{\begin{array}{l} {\varphi _{k,2} =\chi _{k,2} e^{i\omega _{k} \tau_0} } \\ {\varphi _{N-k,2} =\chi _{N-k,2} e^{i\omega_k \tau_0} } \end{array}\right. ,
\] 

and

\[
\left\{\begin{array}{l} {i\frac{\partial \chi _{k,1} }{\partial \tau _{1} } =0} \\ {i\frac{\partial \chi _{N-k,1} }{\partial \tau _{1} } =0} \end{array} \right.
 \] 
The amplitudes $\chi_{k,1}$, $\chi_{N-k,1}$, and $\chi_{k,2}$, and $\chi_{N-k,2}$ do not depend on the times $\tau_0$ and $\tau_1$.
Now we can integrate the equations \eqref{_B4_} with respect to the time $\tau_0$. The requirement of absence of secular terms means that only the terms with zero-value argument of exponent make a contribution in the sum in Eqs \eqref{_B4_}. Finally, after the integration of obtained equations  with respect to the time $\tau_1$ we get:

\begin{equation} \label{_B5_}
\begin{array}{l} {i\frac{\partial \chi _{\frac{N}{2},1} }{\partial \tau _{2} } +\frac{3\beta }{8} \Omega \sum _{l,m,n=1}^{N-1}C_{\frac{N}{2},l,m,n} \chi _{l,1} \chi _{m,1} \chi _{n,1}^{*} \delta (\omega _{l} +\omega _{m} -\omega _{n} -\Omega )   =0} \\ {i\frac{\partial \chi _{\frac{N}{2}\pm 1,1} }{\partial \tau _{2} } -\frac{\pi ^{2} }{2} \Omega \chi _{N/2\pm 1,1} +\frac{3\beta }{8} \Omega  \sum _{l,m,n=1}^{N-1}C_{\frac{N}{2}\pm 1,l,m,n}    \chi _{l,1} \chi _{m,1} \chi _{n,1}^{*} \delta (\omega _{l} +\omega _{m} -\omega _{n} -\Omega )=0} \end{array}
\end{equation} 

It is necessary to note that the frequency spectrum of a finite FPU-system is essentially discrete one and no exact resonances exist except those for the degenerate modes with indices $k$, $N-k$. Therefore only approximate resonant conditions can be satisfied for the highest-frequency modes, as it has been pointed in the Introduction. Taking into account these rules we find the final form of Eqs \eqref{_B5_} (we omit the second index in the subscription of main-order amplitudes $\chi_k$):

\begin{equation} \label{_B6_}
\begin{array}{l} 
{i\frac{\partial \chi _{\frac{N}{2}} }{\partial \tau _{2} } +\frac{3\beta \Omega  }{4}  [(\sum _{l=1}^{N-1}|\chi _{l} |^{2}  +|\chi _{\frac{N}{2}+1} |^{2} +|\chi _{\frac{N}{2}-1} |^{2} )\chi _{\frac{N}{2}} +(\chi _{\frac{N}{2}+1}^{2} +\chi _{\frac{N}{2}-1}^{2} )\chi _{\frac{N}{2}}^{*} ]=0}\\

 {i\frac{\partial \chi _{\frac{N}{2}+1} }{\partial \tau _{2} } -\frac{\pi ^{2} \Omega }{2} \chi _{\frac{N}{2}+1} +\frac{3\beta \Omega }{8} \{ 2\sum _{l=1}^{N-1}|\chi _{l} |^{2} \chi _{\frac{N}{2} +1}  +[(2|\chi _{\frac{N}{2}} |^{2} +|\chi _{\frac{N}{2} +1} |^{2} +} \\
  {|\chi _{N-1} |^{2} -|\chi _{1} |^{2} )\chi _{\frac{N}{2} +1} +(2\chi _{\frac{N}{2}}^{2} +\chi _{\frac{N}{2} -1}^{2} )\chi _{\frac{N}{2} +1}^{*} -(\chi _{1} \chi _{N-1}^{*} +\chi _{N-1} \chi _{1}^{*} )\chi _{\frac{N}{2} -1} ]\} =0} \\ 

{i\frac{\partial \chi _{\frac{N}{2}-1} }{\partial \tau _{2} } -\frac{\pi ^{2} \Omega }{2} \chi _{\frac{N}{2}-1} +\frac{3\beta \Omega }{8}\{ 2\sum _{l=1}^{N-1}|\chi _{l} |^{2} \chi _{\frac{N}{2} -1}  +[(2|\chi _{\frac{N}{2}} |^{2} +|\chi _{\frac{N}{2} -1} |^{2} -} \\ {|\chi _{N-1} |^{2} +|\chi _{1} |^{2} )\chi _{\frac{N}{2} -1} +(2\chi _{\frac{N}{2}}^{2} +\chi _{\frac{N}{2} +1}^{2} )\chi _{\frac{N}{2} -1}^{*} -(\chi _{1} \chi _{N-1}^{*} +\chi _{N-1} \chi _{1}^{*} )\chi _{\frac{N}{2} +1} ]\} =0}. 
\end{array} 
\end{equation} 

It is easy to show that the dynamics of the low-frequency modes $\chi_1$ and $\chi_{N-1}$ is defined by the next equations:

\begin{equation} \label{_B7_}
\begin{array}{l}
 {i\frac{\partial \chi _{1} }{\partial \tau _{2} } +\frac{3\beta \omega _{1} }{8}  [ (\sum _{l=1}^{N-1}|\chi _{l} |^{2} +|\chi _{1} |^{2}) \chi _{1} +\chi _{N-1}^{2} \chi _{1}^{*} +(|\chi _{N/2+1} |^{2} -|\chi _{N/2-1} |^{2} )\chi _{1} -} \\ {\quad \quad \quad (\chi _{N/2-1} \chi _{N/2+1}^{*} +\chi _{N/2+1} \chi _{N/2-1}^{*} )\chi _{N-1} ]=0} \\
  {i\frac{\partial \chi _{N-1} }{\partial \tau _{2} } +\frac{3\beta \omega _{1}  }{8}[ (\sum _{l=1}^{N-1}|\chi _{l} |^{2}+|\chi _{N-1} |^{2}) \chi _{N-1} +\chi _{1}^{2} \chi _{N-1}^{*} +(|\chi _{N/2-1} |^{2} -|\chi _{N/2+1} |^{2} )\chi _{N-1} -} \\ {\quad \quad \quad \quad (\chi _{N/2-1} \chi _{N/2+1}^{*} +\chi _{N/2+1} \chi _{N/2-1}^{*} )\chi _{1} ]=0}
 \end{array}
 \end{equation}

First of all we can see that Eqs \eqref{_B6_}, \eqref{_B7_} couple the high- and low-frequency modes with the pairs of indices ($k, N-k$) and ($N/2-k, N/2+k$). This result is in accordance with the resonant normal form of FPU chain with even number of the particles \cite{r15}.  Moreover one can find that Eqs \eqref{_B6_},  \eqref{_B7_} posses the integrals of motion which are an additional ones to the integrals of energy end momentum:

\[
\begin{array} {l}
 { X_{\frac{N}{2}}=|\chi_{\frac{N}{2}}|^2+|\chi_{\frac{N}{2}+1}|^2+|\chi_{\frac{N}{2}-1}|^2 } \\
 { X_{1}=|\chi_{1}|^2+|\chi_{N-1}|^2 }
\end{array}
\] 

The existence of these integrals allows us to conclude that the sum over all the modes is a constant: $\sum _{l=1}^{N-1}|\chi _{l} |^{2}=\sum _{l=1}^{N/2} X_l=const$. The presence of such terms in Eqs \eqref{_B7_} leads to the frequency shift $\delta\omega_k=(3 \beta \omega_k/4) \sum _{l=1}^{N/2} X_l$. In such a case we can redefine the amplitudes: $\chi_k \rightarrow \chi_k e^{i\delta \omega_k \tau_2}$, and extract the respective terms from the equations.

\end{document}